\documentstyle[12pt]{article}
\def\journal{\topmargin .3in    \oddsidemargin .5in
        \headheight 0pt \headsep 0pt
        \textwidth 5.625in 
        \textheight 8.25in 
        \marginparwidth 1.5in
        \parindent 2em
        \parskip .5ex plus .1ex         \jot = 1.5ex}
\journal

\catcode`\@=11
\def\marginnote#1{}
\catcode`\@=11
\def\section{\@startsection {section}{1}{0pt}{-3.5ex plus -1ex minus
 -.2ex}{2.3ex plus .2ex}{\raggedright\large\bf}}
\catcode`\@=12
\newskip\humongous \humongous=0pt plus 1000pt minus 1000pt

\newif\ifdtup

\def\Q{{\mathchoice
{\setbox0=\hbox{$\displaystyle\rm Q$}\hbox{\raise 0.15\ht0\hbox to0pt
{\kern0.4\wd0\vrule height0.8\ht0\hss}\box0}}
{\setbox0=\hbox{$\textstyle\rm Q$}\hbox{\raise 0.15\ht0\hbox to0pt
{\kern0.4\wd0\vrule height0.8\ht0\hss}\box0}}
{\setbox0=\hbox{$\scriptstyle\rm Q$}\hbox{\raise 0.15\ht0\hbox to0pt
{\kern0.4\wd0\vrule height0.7\ht0\hss}\box0}}
{\setbox0=\hbox{$\scriptscriptstyle\rm Q$}\hbox{\raise 0.15\ht0\hbox to0pt
{\kern0.4\wd0\vrule height0.7\ht0\hss}\box0}}}}
\def\C{{\mathchoice
{\setbox0=\hbox{$\displaystyle\rm C$}\hbox{\hbox to0pt
{\kern0.4\wd0\vrule height0.9\ht0\hss}\box0}}
{\setbox0=\hbox{$\textstyle\rm C$}\hbox{\hbox to0pt
{\kern0.4\wd0\vrule height0.9\ht0\hss}\box0}}
{\setbox0=\hbox{$\scriptstyle\rm C$}\hbox{\hbox to0pt
{\kern0.4\wd0\vrule height0.9\ht0\hss}\box0}}
{\setbox0=\hbox{$\scriptscriptstyle\rm C$}\hbox{\hbox to0pt
{\kern0.4\wd0\vrule height0.9\ht0\hss}\box0}}}}

\font\fivesans=cmss10 at 4.61pt
\font\sevensans=cmss10 at 6.81pt
\font\tensans=cmss10
\newfam\sansfam
\textfont\sansfam=\tensans\scriptfont\sansfam=\sevensans\scriptscriptfont
\sansfam=\fivesans
\def\sans{\fam\sansfam\tensans}
\def\Z{{\mathchoice
{\hbox{$\sans\textstyle Z\kern-0.4em Z$}}
{\hbox{$\sans\textstyle Z\kern-0.4em Z$}}
{\hbox{$\sans\scriptstyle Z\kern-0.3em Z$}}
{\hbox{$\sans\scriptscriptstyle Z\kern-0.2em Z$}}}}

\mathchardef\endbar="375

\def\ceilfill{$\raise3pt\hbox{$\mathsurround=0pt\mathord\endbar$}
  \mkern-2mu \xleaders\hbox{$\mkern-5mu
  \mathord-\mkern-5mu$}\hfill\mkern-7mu
  \raise3pt\hbox{$\mathsurround=0pt\mathord\endbar$}$}

\def\floorfill{$\raise9pt\hbox{$\mathsurround=0pt\mathord\endbar$}
  \mkern-2mu \xleaders\hbox{$\mkern-5mu
  \mathord-\mkern-5mu$}\hfill\mkern-7mu
  \raise9pt\hbox{$\mathsurround=0pt\mathord\endbar$}$}

\def\overcontract#1{\mathop{\vbox{\ialign{##\crcr\noalign{\kern3pt}
  \ceilfill\hskip6pt\crcr\noalign{\kern3pt\nointerlineskip}
  $\hfil\displaystyle{#1}\hfil$\crcr}}}}

\def\undercontract#1{\mathop{\vtop{\ialign{##\crcr
  $\hfil\displaystyle{#1}\hfil$\crcr\noalign{\kern3pt\nointerlineskip}
  \floorfill\hskip6pt\crcr\noalign{\kern3pt}}}}}
\def\a{\alpha}
\def\b{\beta}

\def\d{\delta}

\def\m{\mu}
\def\n{\nu}
\def\t{\tau}
\def\p{\pi}

\def\o{\omega}

\def\et{\eta}

\def\D{\Delta}
\def\bz{\bar{z}}

\def\tL{\tilde{L}}
\def\tT{\tilde{T}}

\def\tz{\tilde{z}}

\def\F{{\cal F}}

\def\T{{\cal T}}
\def\C{{\cal C}}
\def\cO{{\cal O}}

\def\limit#1#2{ \smash{ \mathop{#1} \limits_{#2} } }
\def\rank{{\rm rank\,}}

\def\pa{\partial}

\def\tpa{\tilde{\partial}}
\def\ve{\vert}

\def\ra{\rightarrow}

\def\ci{\cite}
\def\xx{\hbox{ }^*_*}

\def\ref#1{$^{#1)}$}

\begin{document}
\begin{titlepage}
\begin{center}
 November, 1994        \hfill     UCB-PTH-94/33 \\
hep-th/9411209    \hfill     LBL-36437 \\
\vskip .5in

{\large \bf Irrational Conformal Field Theory \\ on the Sphere and the Torus}
\footnote{This work was supported in part by the Director, Office of
Energy Research, Office of High Energy and Nuclear Physics, Division of
High Energy Physics of the U.S. Department of Energy under Contract
DE-AC03-76SF00098 and in part by the National Science Foundation under
grant PHY90-21139.}
\vskip .3in
M.B. Halpern
\footnote{e-mail: MBHALPERN@LBL.GOV}
\footnote{Invited talk at the XIth International Congress of Mathematical
Physics, Paris, July 18-23, 1994.}
\vskip .2in
{\em  Department of Physics \\ University of California\\ and \\
      Theoretical Physics Group \\ Physics Division \\
     Lawrence Berkeley Laboratory\\ 1 Cyclotron Road \\
      Berkeley, California 94720 \\
       USA}

\end{center}
\vskip .3in
\begin{abstract}
I review the foundations of irrational conformal field theory (ICFT),
which includes rational conformal field theory as a small subspace.
Highlights of the review include the Virasoro master equation and the
generalized Knizhnik-Zamolodchikov equations for the correlators of
ICFT on the sphere and the torus.

\end{abstract}

\end{titlepage}
\renewcommand{\thepage}{\arabic{page}}
\setcounter{page}{1}
\setcounter{footnote}{1}

\setcounter{section}{-1}
\section{Outline of the Talk}
\begin{enumerate}
\item History of the affine-Virasoro constructions.
\item The general affine-Virasoro construction.
\item Irrational conformal field theory.
\item The Ward identities of ICFT.
\item The generalized KZ equations of ICFT.
\item ICFT on the torus.
\item Conclusions.
\end{enumerate}

\section{History of the Affine-Virasoro Constructions}
 Affine-Virasoro constructions are Virasoro operators constructed with the
currents $J_a,\, a=1 , \ldots,\mbox{dim}\,g$ of affine Lie $g$. All known
conformal field theories may be constructed this way.

Here is a brief history of these constructions, which began with two\break
papers~[1,2]
by Bardakci and myself in 1971. These papers contained the following
developments.

\noindent a) The first examples of affine Lie algebra, or current algebra on
$S^1$. This was the independent discovery of affine Lie algebra in physics,
including the
\hbox{affine central extension some years before it was
recognized in mathematics~\ci{lw}.}
 Following the convention in math,
we prefer the functional names affine
Lie
algebra or
current algebra, reserving ``Kac-Moody'' for the more general case~\ci{km}
including hyperbolic algebras.

\noindent b) World-sheet fermions (half-integer moded), from which the affine
algebras were constructed. Ramond \ci{ram}
gave the integer-moded case in the same issue of Physical Review.

\noindent c) The first affine-Sugawara constructions, on the currents of affine
Lie algebra. Sugawara's model \ci{sug} was in four dimensions on a
different algebra. The
affine-Sugawara
constructions were later generalized by Knizhnik and
Zamolodchikov~\ci{kz}
and
Segal~\cite{se}, and the corresponding world-sheet action was given
by \hbox{Witten}~\ci{wit1}.

\noindent d) The first coset constructions, implicit in [1] and
explicit in [2], which were later generalized by Goddard, Kent and
Olive [10].

What was forgotten for many years was that our first paper also gave another
affine-Virasoro
construction, the so-called ``spin-orbit'' model, which was more general than
the affine-Sugawara and coset constructions.

\section{The General Affine-Virasoro Construction}
Motivated by the affine-Sugawara, coset and spin-orbit constructions, E.
Kiritsis
and I found the general affine-Virasoro construction [11,12] in 1989.

The construction begins with the currents of untwisted affine
Lie $g $  [4,1]
$$J_a(z)\, J_b(w)={G_{ab} \over (z-w)^2}+ { i{f_{ab}}^c \over z-w} \, J_c(w) +
\cO (z-w)^0 \eqno(2.1) $$
where $a,b = 1,\ldots, \mbox{dim}\,g$ and ${f_{ab}}^c$ and $G_{ab}$ are
the structure constants and
generalized Killing metric of
$g= \oplus_I g_I $. To obtain level
$x_I = 2k_I/\psi_I^2$ of affine $g_I$ with dual Coxeter number
$\tilde{h}_I= Q_I /\psi_I^2$, take
$$G_{ab}=\oplus_I k_I \, \et_{ab}^I\;\;\;\;,\;\;\;\;
f_{ac}{}^d f_{bd}{}^c = - \oplus_I Q_I \, \eta_{ab}^I
\eqno(2.2) $$
where $\eta_{ab}^I$ and $\psi_I$ are the Killing metric and
 highest root of $g_I$.

Given the currents, we consider the general stress tensor
$$ T= L^{ab} \xx J_a J_b \xx + D^a \partial J_a + d^a J_a
\eqno(2.3) $$
where the coefficients $L^{ab}= L^{ba}$, $D^a$ and $d^a$ are to be determined.
$L^{ab}$ is called the inverse inertia tensor, in analogy with the spinning
top. The stress tensor is required to satisfy the Virasoro algebra
$$T(z)\,T(w)={c/2 \over (z-w)^4} +\left(\frac{2}{(z-w)^2}+{ \pa_w \over z-w}
\right)T(w)+ \cO (z-w)^0 \eqno(2.4)$$
where $c$ is the central charge. I give here only the result at $D^a=d^a=0$,
which is called the Virasoro master equation [11,12]
$$ L^{ab}=2 L^{ac}G_{cd}L^{db}-L^{cd}L^{ef}{f_{ce}}^{a}{f_{df}}^{b}-
L^{cd}{f_{ce}}^{f}{f_{df}}^{(a}L^{b)e} \eqno(2.5a)$$
$$ c=2 \,G_{ab}L^{ab}\;\;\;\;.\eqno(2.5b)$$
The master equation has been identified \ci{hy1} as an Einstein
system on the group manifold, with
$$ g^{ij}= e_a^i L^{ab} e_b^j \;\;\;\;,\;\;\;\;
c=\mbox{dim}\,g-4R \eqno(2.6)$$
where $g^{ij}$ and $R$  are the inverse Einstein metric
and  Einstein curvature scalar respectively.

Here are some simple facts about the master equation which we will need in
this lecture.

\noindent a) Affine-Sugawara constructions [1,2,7-9]. The
affine-Sugawara construction on $g$ is
$$ L_g^{ab}=\oplus_I {\eta_I^{ab} \over 2k_I +Q_I}\;\;\;\,,\;\;\;\;\;\;\;
c_g=\sum_I {x_I \, \mbox{dim}\,g_I \over x_I + \tilde{h}_I } \eqno(2.7) $$
and similarly for $L_h$ on $h \subset g$.

\noindent b) K-conjugation covariance [1,2,10,11]. Solutions of the
master equation come in commuting K-conjugate pairs
$T=L^{ab} \xx J_a J_b \xx$ and
$\tilde{T}=\tilde{L}^{ab} \xx J_a J_b \xx$, which sum to the affine-Sugawara
construction $T_g=L_g^{ab} \xx J_a J_b \xx$,
$$ L + \tilde{L}=L_g \;\;\;,\;\;\;\;\;\;\; T + \tilde{T} = T_g
\;\;\;\;,\;\;\;\;
c + \tilde{c}=c_g
\eqno(2.8a) $$
$$ T(z)\, \tilde{T} (w)= \cO (z-w)^0  \;\;\;\;\;\; . \eqno(2.8b) $$
K-conjugation is the central feature of affine-Virasoro constructions, and it
suggests that the affine-Sugawara construction should be thought of as the
tensor product of any pair of K-conjugate conformal field theories.
This is the conceptual basis of factorization, discussed in Sections 4-7.

\noindent c) Coset constructions [1,2,10].
K-conjugation generates new solutions from
old. The simplest examples are the $g/h$ coset constructions
$$ \tilde{L} =  L_g - L_h = L_{g/h} \;\;\;,\;\;\;\;
\tilde{T} =  T_g - T_h = T_{g/h}
\;\;\;,\;\;\;\; \tilde{c} = c_g-c_h = c_{g/h} \eqno(2.9)$$
which follow by K-conjugation from $L_h$ on $h \subset g$.
Repeated K-conjugation on the nested subalgebras
$g\supset h_1 \cdots \supset h_n$ gives the affine-Sugawara nests [14,15],
which have been shown to be tensor product theories formed by
tensoring the conformal blocks of appropriate subgroups and nests [39,43].

\noindent d) High-level expansion.
At high-level on simple $g$, the leading behavior of the inverse inertia
tensors is [21,29]
$$ \tilde L^{ab}={\tilde P^{ab} \over 2k} + \cO(k^{-2})
\quad , \quad L^{ab}={L^{ab}\over 2k} + \cO(k^{-2})
\eqno(2.10a) $$
$$ \tilde L^{ab}+L^{ab}=L_g^{ab}={\eta^{ab}\over 2k} + \cO(k^{-2})
\eqno(2.10b) $$
where $\tilde P$ and $P$ are the high-level projectors of the $\tilde L$
and the $L$ theory respectively.
Higher-order terms in this expansion are studied in Ref.\ci{hl}.

\section{Irrational Conformal Field Theory}
Here is an overview of the solution space of the master equation, called
affine-Virasoro space. For further details see the review in Ref.\ci{rd}

\noindent a) Counting [15,18]. The master equation is a set of
$\mbox{dim}\,g (\mbox{dim}\, g + 1)/2$ coupled quadratic equations on the same
number of unknowns $L^{ab}$. This allows us to estimate the number of
inequivalent solutions on each manifold. As an example, there are
approximately $\frac{1}{4}$ billion conformal field theories on each level
of affine $SU(3)$, and exponentially larger numbers on larger manifolds.

\noindent b) Exact solutions
[15,18-26,16].
Large numbers of new solutions have been
found in closed form. On positive integer levels of affine compact $g$, most of
these
solutions are unitary with irrational central charge. As an example, the value
at level 5 of affine $SU(3)$ \ci{hl}
$$ c \left( (SU(3)_5)_{D(1)}^{\#} \right) = 2 \left( 1 - {1 \over \sqrt{61}}
\right) \simeq 1.7439 \eqno(3.1) $$
is the lowest unitary irrational central charge yet observed. See Ref.\ci{lie}
for the most recent list of exact solutions.

\noindent c) Systematics \ci{rd}. Generically, affine-Virasoro space is
organized into level-families of conformal field theories, which are
essentially
analytic functions of the level. On positive integer level of affine compact
$g$, it is clear from the form of the master equation that the generic
level-family has generically irrational central charge.
Since rational central charge is sporadic in affine-Virasoro space,
we refer to the space of all CFT's as irrational conformal field theory
(ICFT),
$$ICFT \supset\supset RCFT.
\eqno(3.2) $$
Moreover, rational central charge is rare in the subspace of unitary conformal
field theories. Indeed, the rational conformal field theories are rare in the
space of Lie $h$-invariant conformal field theories \ci{lie}, which are
themselves
quite rare (see Section 6).
Many candidates for new rational conformal field theories
[23,27,28], beyond the coset constructions, have also been found.

\noindent d) Classification\footnote{In the course of this work, a new
and apparently fundamental connection between Lie groups and graphs was seen.
The interested reader should consult Ref.\ci{lie} and especially Ref.\ci{ggt},
which axiomatizes these observations.}. Study of the space by
high-level
expansion \ci{hl} shows a partial classification by graph theory and
generalized graph theories  [18,22-26].
In the classification,
the high-level projectors $\tilde P$ and $P$ in eq.(2.10)
are the
adjacency matrices of the graphs, and each graph is a
level-family, whose conformal field theories carry the symmetry of the graph.
At the present time, seven graph-theory units \ci{rd} of generically unitary
and
irrational conformal field theories have been studied, and it seems likely
that many more can be found \ci{ggt}.

Large as they are, the graph theories classify only very small regions of
affine-Virasoro space. Enough has been learned however to see that all known
exact solutions are special cases of relatively high symmetry, whereas the
generic conformal field theory is completely asymmetric \ci{gt}.

Before going on to the Ward identities, I should mention several other lines
of development.

\noindent 1. Non-chiral CFT's. Adding right-mover copies $\bar{T}$ and
$\tilde{\bar{T}}$ of the K-conjugate stress tensors, we may take the usual
Hamiltonian $H= L^{(0)} + \bar{L}^{(0)}$ for the $L$ theory. Because
the generic CFT has no residual affine symmetry, the physical Hilbert space
of the generic theory $L$ is characterized by
$$ \tilde{L}{(m>0)} |L \mbox{-physical} \rangle =
\tilde{\bar{L}}{(m>0)} |L \mbox{-physical} \rangle =0 \;\;\;\;
\eqno(3.3) $$
where $\tilde L{(m)}$ and $L{(m)}$ are the modes of $\tilde T$ and $T$.

\noindent 2. World sheet action [29,42]. Correspondingly, the world-sheet
action  of the generic theory
$L$ is a spin-two gauge theory, in which the WZW theory is gauged by the
K-conjugate theory $\tL$. An open direction here is the relation with
$\sigma$-models and the corresponding space-time geometry of
irrational conformal field theory. In this connection, see also
Ref.\ci{hy1}.

\noindent 3. Superconformal master equations \ci{sme}. The N=1 system has been
studied in some detail [23-26],
and a simplification \ci{jos} of the N=2
system has recently been noted: Eq.(D.5) of Ref.\ci{sme} is redundant, so that
eq.(D.6) is the complete N=2 system. It is an important open problem to find
unitary  N=2 solutions with irrational central charge.
Another open question at N=2 is the relation of the master equation to the
bosonic constructions of Kazama and Suzuki [27,28].

\noindent 4. Exact C-functions [32,30,17]. Exact C-functions \ci{zam}
and associated C-theorems are known for the N=0 and N=1 master equations, but
not yet for N=2.

\section{The Ward Identities of ICFT}
It is clear that the Virasoro master equation is the first step in the study
of irrational conformal field theory, which contains rational conformal
field theory as a small subspace of relatively high symmetry.
On the other hand, the correlators of irrational conformal field theory have,
until recently, remained elusive. The reason is that most of the computational
methods of conformal field theory have been based on null states of extended
Virasoro algebras [5,34-36], which apparently have little to say about the
affine-Sugawara constructions and the general ICFT.

Recently, N. Obers and I have found a set of equations, the affine-Virasoro
Ward identities \ci{co}, for the correlators of irrational conformal field
theory.

The first step is to consider KZ-type null states [7,37],
which live in the modules
of the universal covering algebra of the affine algebra, and which are more
general than extended Virasoro null states. As an example, we have
$$ 0 = L{(-1)} | R_g \rangle - 2 L^{ab} J_a{(-1)} | R_g \rangle \T_b
\eqno(4.1) $$
which holds for all affine-Virasoro constructions $L^{ab}$. Here
$|R_g \rangle^{\a}, \; \a =1,\ldots ,\mbox{dim}\,\T $ is the affine primary
state with matrix representation $\T$, and the original KZ null state is
recovered from (4.1) by taking $L^{ab}=L_g^{ab} $ and $L{(-1)}=  L_g{(-1)}$.
We also choose a so-called $L$-basis of representation $\T$, in which the
conformal weight matrix is diagonal \ci{co}
$$  L^{ab} {(\T_a \T_b)_{\a}}^{\b} = \D_{\a}(\T) \, \d_{\a}^{\b}
\;\;\;\;,\;\;\;\; \a,\b =1 , \ldots ,\mbox{dim}\,\T \;\;\;\; . \eqno(4.2a) $$
$$ L({m \geq 0}) | R_g \rangle =\d_{m,0} \, \D_{\a}(\T)
| R_g \rangle \;\;\;\;,\;\;\;\;
 \tilde{L}({m \geq 0}) | R_g \rangle =\d_{m,0} \,
\tilde{\D}_{\a}(\T)  | R_g \rangle \eqno(4.2b) $$
$$  \D_{g}(\T) =  \D_{\a}(\T) + \tilde{\D}_{\a}(\T) \;\;\;\;. \eqno(4.2c) $$
In such a basis, the states $|R_g \rangle$ are called the broken affine primary
states, because the conformal weights $\D_g(\T)$ of $|R_g \rangle$ under
the affine-Sugawara
construction are broken to the conformal weights
$\D_{\a} (\T),\tilde\D_\a(\T)$ of the
$L$ theory and the $\tilde L$ theory.
These states are also examples of Virasoro biprimary states, which
are simultaneously Virasoro primary under both of the commuting K-conjugate
stress tensors.

{}To use these null states in correlators, we need the
Virasoro primary fields of the $L$ and the $\tilde L$ theories.
Because we have two commuting
stress tensors, the natural objects are the Virasoro biprimary fields [38,37]
$$ R^{\a}(\tz,z) =\mbox{e}^{(\tz-z) \tL{(-1)} } R_g^{\a} (z) \,
\mbox{e}^{(z-\tz) \tL{(-1)}} = \mbox{e}^{(z-\tz) L{(-1)} } R_g^{\a} (\tz) \,
\mbox{e}^{(\tz-z) L{(-1)}} \eqno (4.3a) $$
$$ R(z,z) = R_g(z) \;\;\;\;,\;\;\;\;
 R (0,0)  |0 \rangle= | R_g \rangle  \eqno(4.3b) $$
which are  simultaneously Virasoro primary under $T$ and $\tT$.
In (4.3), $R_g^{\a}$ are the broken affine primary fields, and the complex
variables $\tz,z$ are independent. The averages
of these bilocal fields
$$ A^{\a}(\tz,z) =  \langle R^{\a_1}(\T^1,\tz_1,z_1) \ldots
R^{\a_n}(\T^n,\tz_n,z_n) \rangle
\eqno(4.4) $$
are called biconformal correlators or bicorrelators.

Inserting the null state (4.1) into the bicorrelator, the Virasoro
term gives derivatives with respect to the $z$'s, as usual, while the current
term can be evaluated, in terms of the representation matrices, on the
affine-Sugawara line $\tz =z$. More generally, we obtain the
affine-Virasoro Ward identities \ci{co},
$$   \tpa_{j_1} \ldots \tpa_{j_q} \pa_{i_1} \ldots \pa_{i_p} A^{\a}(\tz,z)
 \ve_{\bz=z} = A^{\b}_g (z) { W_{j_1 \ldots j_q, i_1 \ldots i_p }
(z)_{\b}}^{\a} \eqno(4.5) $$
where $W_{ j_1 \ldots j_q ,i_1 \ldots i_p} $ are the affine-Virasoro
connection moments
and $A(z,z) = A_g(z)$ is the affine-Sugawara correlator (which
satisfies the KZ equation on $g$). The first-order connection moments are
$$  W_{0,i} = 2 L^{ab} \sum_{j\neq i}^n
{\T_a^i \T_b^j \over z_{ij} }  \;\;\;\;,\;\;\;\;\;
W_{i,0} = 2 \tL^{ab} \sum_{j\neq i}^n  { \T_a^i \T_b^j \over z_{ij} }
 \eqno(4.6) $$
which follow from the null state (4.1) and its K-conjugate copy with
$L \ra \tL$. The sum of these two connections is the KZ connection
$W_i^g$, with $L=L_g$.

All the connection moments
may be computed by standard dispersive techniques from the
formula [37,39]
$$  A_g^{\b} (z) {W_{j_1 \ldots j_q, i_1 \ldots i_p } (z)_{\b}}^{\a} =
\;\; \;\;\;\;\;\;\;\;\;\;\; \;\;\;\;\;\;\;\;\;\; $$
$$ \left[ \prod_{r=1}^{q} \tL^{a_r b_r} \oint_{z_{j_r}}
{\mbox{d}\o_r \over 2\p i} \oint_{\o_r}  {\mbox{d}\et_r \over 2\p i} \;
{1 \over \et_r-\o_r} \right] \!
\left[ \prod_{s=1}^{p} L^{c_s d_s} \oint_{z_{i_s}} \!
{\mbox{d}\o_{q+s} \over 2\p i} \oint_{\o_{q+s}} \!
{\mbox{d}\et_{q+s} \over 2\p i} \; {1 \over \et_{q+s}-\o_{q+s}} \right] $$
$$ \times  \langle
J_{a_1}(\et_1)  J_{b_1}(\o_1)  \ldots J_{a_q}(\et_q)  J_{b_q}(\o_q)
J_{c_1}(\et_{q+1})  J_{d_1}(\o_{q+1})  \ldots  $$
$$ \;\;\;\; \;\;\;\;\;\; \; J_{c_p}(\et_{q+p})  J_{d_p}(\o_{q+p})
R_g^{\a_1} (\T^1,z_1) \ldots  R_g^{\a_n} (\T^n,z_n) \rangle
 \eqno(4.7) $$
since the required averages are in the affine-Sugawara theory on $g$. The
results (4.5) and (4.7) prove the existence of the biconformal correlators
(at least
as an expansion about the affine-Sugawara line $\tz =z$), but computation of
all the connections appears to be a formidable task. So far, we have explicitly
evaluated the first and second-order connection moments
\ci{co} for all theories,
and all the connection moments for the coset constructions [37,39] and
affine-Sugawara nests \ci{wi}.
We have also found the leading term of the connection moments for the
general ICFT at high level on simple $g$ \ci{wi}.

\section{The Generalized KZ equations of ICFT.}

The Ward identities (4.5) may be written as generalized KZ equations \ci{flat}
$$ \tpa_i A(\tz,z)=A(\tz,z)\tilde W_i(\tz,z) \quad , \quad
\pa_i A(\tz,z)=A(\tz,z)W_i(\tz,z)
\eqno(5.1a) $$
$$ A(z,z)=A_g(z) \eqno(5.1b) $$
where $\tilde W_i, W_i$ are flat connections and (5.1b) is the
affine-Sugawara boundary condition for the system.  The flat connections can
be written in terms of the connection moments via the non-linear relations,
$$ \tilde W_i=\tilde F^{-1}\tpa_i \tilde F \quad , \quad
W_i=F^{-1}\pa_i F
\eqno(5.2a) $$
$$ \tilde F(\tz,z) = \sum_{q=0}^{\infty} {1 \over q !} \sum_{j_1 \ldots j_q}
\prod_{\n=1}^q (\tz_{j_\n} - z_{j_\n } ) W_{j_1 \ldots j_q,0} (z)
\eqno(5.2b) $$
$$  F(\tz,z) = \sum_{p=0}^{\infty} {1 \over p !} \sum_{i_i \ldots i_p}
\prod_{\m=1}^p (z_{i_\m} - \tz_{i_\m } ) W_{0,i_1 \ldots i_p} (\tz)
\eqno(5.2c) $$
where $\tilde F$ and $F$ are the evolution operators of the flat connections.

The flat connections satisfy
$$ \tilde W_i(z,z)+W_i(z,z) =W_i^g(z)=2L^{ab}_g\sum_{j\ne i}
{\T_a^i \T_b^j \over z_i-z_j}
\eqno(5.3)
$$
where $W_i^g$ is the affine-Sugawara connection which appears in the
KZ equation \ci{kz}
$$\pa_i A_g=A_g W_i^g. \eqno(5.4) $$
Indeed, the KZ equation is implied by eq.(5.3) and the generalized KZ
equations (5.1),
$$\pa_i A_g(z)=(\tpa_i+\pa_i)A(\tz,z)|_{\tz=z} = A_g(z)W_i^g(z).
\eqno(5.5) $$
where the second step is a chain rule.  Moreover, the generalized KZ
equations include the KZ equation itself as a special case when we
choose the simplest K-conjugate pair $\tilde L=0, L=L_g$.
The resulting system in this case,
$$ \pa_i A=0 \quad , \quad \pa_i A=A W_i^g
\eqno(5.6) $$
is equivalent to (5.4).

The generalized KZ equations have been solved exactly for the coset
constructions [37,39,41], the affine-Sugawara nests [39,41] and the
general ICFT at high-level on simple $g$ [39,41].
In all these cases, one sees a factorization of the bicorrelators
$$ A(\tz,z)=\sum_\n \tilde A_\n(\tz)A_\n(z)
\eqno(5.7)
$$
into the correlators of the individual conformal field theories
$\tilde L$ and $L$.

I mention in particular the results for the four-point correlators of the
coset constructions, whose conformal blocks, called the coset blocks
\ci{co},
$$\C_{g/h}=\F_g\F_h^{-1}
\eqno(5.8) $$
turn out to be the conformal blocks conjectured for the coset constructions
by Douglas \ci{do}.  Here $\F_g$ and $\F_h$ are matrices of conformal
blocks of the affine-Sugawara constructions on $g$ and on $h$.
See \ci{co} for a detailed example of coset blocks on
$(SU(n)\times SU(n))/SU(n)$.

I also mention the candidate correlators \ci{wi} obtained by
factorization for all ICFT.  Here, the factorization (5.7) is studied
assuming the flat connections of the general theory are given as input
data. In the general case the factorization is not unique, but a
factorization has been found whose candidate correlators so far appear
to be completely physical.  In particular, the correlators reproduce the
known correlators of the coset constructions, the affine-Sugawara nests
and the high-level correlators of the general ICFT on simple $g$. For the
general ICFT at finite level, it appears that one obtains an infinite
number of conformal blocks, in accord with intuitive notions about ICFT.
Finally, the correlators exhibit a universal braiding across all ICFT
which includes and generalizes the Fuchsian braiding of RCFT. A review
of this development is given in Ref.\ci{rp}.

\section{ICFT on the Torus}

This year, N. Sochen and I \ci{hs} have extended these developments to
begin the study of ICFT on
the torus.

For each K-conjugate pair $\tilde T, T$ of affine-Virasoro constructions
on integer level of affine compact $g$, the biconformal characters or
bicharacters are defined as
$$
\chi(\T,\tilde\tau,\tau,h)=
Tr_{\T}\left(\tilde q^{\tilde L(0)-\tilde c/24}q^{L(0)-c/24}h\right)
\eqno(6.1a) $$
$$ \tilde q=e^{2\pi i \tilde \tau} \quad , \quad
q=e^{2\pi i \tau}
\eqno(6.1b) $$
where $\tilde L(0),L(0)$ are the zero modes of $\tilde T$ and $T$.  The
source $h$ in (6.1) is an element of a compact Lie group
$H\subset G$ and the trace is over the integrable affine irrep $V_\T$
corresponding to matrix irrep $\T$ of $g$.  The bicharacters satisfy the
affine-Sugawara boundary condition
$$ \chi(\T,\tau,\tau,h)=\chi_g(\T,\tau,h)=
Tr_\T\left(q^{L_g(0)-c_g/24}\right)
\eqno(6.2) $$
where $\chi_g(\T)$ is the conventional
affine-Sugawara character of irrep $V_\T$.

Following the development on the sphere, we found that the
bicharacters satisfy the heat-like differential system,
$$\pa_{\tilde\tau}\chi(\T,\tilde\tau,\tau,h)=
\tilde D(\tilde\tau,\tau,h)\chi(\T,\tilde\tau,\tau,h)
\eqno(6.3a) $$
$$\pa_\tau\chi(\T,\tilde\tau,\tau,h)=
D(\tilde\tau,\tau,h)\chi(\T,\tilde\tau,\tau,h)
\eqno(6.3b) $$
$$\chi(\T,\t,\t,h)=\chi_g(\T,\t,h)
\eqno(6.3c) $$
where $\tilde D$ and $D$ are flat connections.  As discussed in \ci{hs},
this system
includes and generalizes Bernard's heat equation \ci{ber} for
the affine-Sugawara characters.

The heat-like system (6.3) has been solved for the coset constructions
and the general ICFT at high level on simple $g$.

For $h$ and $g/h$, the bicharacters factorize,
$$
\chi_g(\T,\tilde\tau,h)=\sum_{\T^h}\vphantom{\Biggl(}'
\chi_{g/h}(\T,\T^h,\tilde\tau)\chi_h(\T^h,\tilde\tau,h)
\eqno(6.4) $$
where $\chi_h(\T^h)$ are the $\hat h$-characters for $h\subset g$ and the
sum is over all affine irreps $\T^h$ of $\hat h$ at the induced level of
the subalgebra.
The quantities $\chi_{g/h}(\T,\T^h)$ are the coset characters, for
which we obtain a new integral representation,
$$
\chi_{g/h}(\T,\T^h,\tilde\tau)=
\int dh\chi_h^\dagger(\T^h,\tilde\tau,h)\chi_g(\T,\tilde\tau,h) \hfill
\eqno(6.5a)
$$
$$={\tilde q^{-{c_{g/h}\over 24}}\over f(\T^h,\tilde \tau) }
\sum_{\T', \T'^h}
N_{\T'}^{\T}N_{\T'^h}^{\T^h}
\left(\int dh{Tr(h^*(\T'^h))Tr(h(\T'))\over\Pi(\tilde\tau,\sigma(h))}\right)
\tilde q^{\Delta_g(\T')+\Delta_h(\T'^h)}
\eqno(6.5b)
$$
$$f(\T^h,\tau)\equiv\sum_{\T'^h} |N_{\T'^h}^{\T^h}|
q^{2\Delta_h(\T'^h)}
\eqno(6.5c)$$
where $\D_g$ and $\D_h$ are the conformal weights of $\T$ and $\T^h$,
and $dh$ is Haar measure on Lie $h$.
The explicit forms of the coefficients $N$ and
the dual characters $\chi_h^\dagger(\T^h)$
are given in \ci{hs}.
The general result in (6.5a) holds for semisimple $g$ and simple $h$, and
is the analogue on the torus of the formula (5.8) for the coset blocks on
the sphere.  The special case in (6.5b) is the explicit form of (6.5a)
for simple $g$.

For the high-level case, the leading terms of the flat connections are
$$ \tilde D(\tilde L,\tilde\tau,\tau,g)
=2\pi i\left(\sum_{n>0}nTr\big({X_n\over 1-X_n}\tilde P\big)
-{\rank\tilde P\over 24}\right)+\cO(k^{-1})
\eqno(6.6a) $$
$$ D(L,\tilde\tau,\tau,g)=2\pi i\left(\sum_{n>0}nTr\big({X_n\over 1-X_n}P\big)
-{\rank P\over 24}\right)+\cO(k^{-1})
\eqno(6.6b) $$
$$ X_n(\tilde\tau,\tau,g)\equiv (\tilde q^n\tilde P+q^nP)\Omega(g)
\eqno(6.6c) $$
where $\Omega(g)$ is the adjoint action of $g$ and $\tilde P,P$ are the
high-level projection operators (2.10) of the $\tilde L$ and the $L$ theory
respectively.  The result for the high-level bicharacters is given
in \ci{hs}.

So far, factorization of the bicharacters has been studied only for the case
of the Lie $h$-invariant CFT's \ci{lie}, which is the subspace of all ICFT's
with a Lie symmetry $h\subset g$,
$$
\mbox{ICFT}\supset\supset\mbox{Lie $h$-invariant CFT's}\supset\supset
\mbox{RCFT}.
\eqno(6.7) $$
In this case, we have obtained the candidate high-level low-spin
characters of the Lie-$h$ invariant CFT's,
$$\chi_{\tilde L}^{\phantom g}(T,T^h,\tilde\tau) \limit{=}{k}
\int dh
{Tr(h^*(\T^h))Tr(h(\T))
\over
\tilde q^{\rank\tilde P\over 24}
\Pi(\tilde P,\tilde \tau,\Omega(h))}
\eqno(6.8a) $$
$$ \chi_L^{\phantom g}(\T^h,\tau,h)\limit{=}{k}
{Tr(h(\T^h))\over
q^{\rank P\over 24}
\Pi(P,\tau,\Omega(h))}
\eqno(6.8b) $$
where $h(\T)$ and $h(\T^h)$ are the representation matrices of $\T$ and
$\T^h$ in Lie $H$ and low-spin means that the Casimir operator of $T$
is $\cO(k^0)$.
The results (6.8) correctly contain the high-level low-spin characters
of $g/h$ and $h$ but still must be tested for modular covariance, or
further decomposed until modular covariance is obtained.  For
this it will first be necessary to obtain the high-level high-spin
bicharacters, following the method of \ci{hs}.

I finally mention the geometric formulation \ci{hs} of the heat-like
system (6.3) on a source $\hat g$ in the affine Lie group.
In this formulation, one finds a new first-order differential
representation of affine $g\times g$,
$$ E_a(m)=-i e_{am}{}^{i\m}(\pa_{i\m}-ik e_{i\m}{}^y) \quad , \quad
\bar E_a(m)=-i \bar e_{am}{}^{i\m} (\pa_{i\m}+ik\bar e_{i\m}{}^y)
\eqno(6.9)
$$
where the $e$'s and $\bar e$'s are vielbeins and inverse vielbeins on
the affine group manifold, whose coordinates are $y$ (for the central term)
and $x^{i\m}, i=1\ldots\dim g, \m\in\Z$.  The flat connections of the
heat-like system are then obtained in closed form
$$ \tilde D(\hat g)=2\p i \tilde L^{ab}
\left(E_a(0)E_b(0) + 2\sum_{m>0}E_a(-m)E_b(m)\right)
\eqno(6.10a) $$
$$ D(\hat g)=2\p i L^{ab}
\left(E_a(0)E_b(0) + 2\sum_{m>0}E_a(-m)E_b(m)\right)
\eqno(6.10b) $$
as commuting generalized Laplacians on the
affine group. See Ref.\ci{hs} for further discussion of the heat-like
system in this case.

\section{Conclusions}

The Virasoro master equation describes irrational conformal field
theory (ICFT), which includes rational conformal field theory as a small
subspace.  Generalized Knizhnik-Zamolodchikov (KZ) equations have been
derived for the correlators of ICFT on the
sphere and on the torus.
Study of the generalized KZ equations is in an early stage, but so far
has led to new results for the coset constructions and partial results
for all ICFT at high level on simple $g$.

\section*{Acknowledgement}
I am grateful to K. Clubok for his generous help in preparing this
manuscript.


\begin{thebibliography}{99}
\bibitem{bh} K. Bardak\c ci  and M.B. Halpern, Phys. Rev. {\bf D3} (1971) 2493.
\bibitem{h} M.B. Halpern, Phys. Rev. {\bf D4} (1971) 2398.
\bibitem{lw} J. Lepowsky and R.L. Wilson, Comm. Math. Phys. {\bf 62} (1978)
            43.
\bibitem{km}  V.G. Kac, Funct. Anal. App. {\bf 1} (1967) 328;
            R.V. Moody, Bull. Am. Math. Soc. {\bf 73} (1967) 217.
\bibitem{ram} P. Ramond, Phys. Rev. {\bf D3} (1971) 2415.
\bibitem{sug} H. Sugawara, Phys. Rev. {\bf 170} (1968) 1659; C. Sommerfeld,
            Phys. Rev. {\bf 176} (1968) 2019.
\bibitem{kz} V.G. Knizhnik and A.B. Zamolodchikov, Nucl. Phys. {\bf B247}
             (1984) 83.
\bibitem{se}  G. Segal, unpublished.
\bibitem{wit1} E. Witten, Comm. Math. Phys. {\bf 92} (1984) 455;
\bibitem{gko} P. Goddard, A. Kent and D. Olive, Phys. Lett. {\bf B152} (1985)
              88.
\bibitem{hk} M.B. Halpern and E. Kiritsis, Mod. Phys. Lett. {\bf A4} (1989)
1373;
             Erratum ibid. {\bf A4} (1989) 1797.
\bibitem{rus} A.Yu Morozov, A.M. Perelomov, A.A. Rosly, M.A. Shifman and
             A.V. Turbiner, Int. J. Mod. Phys. {\bf A5} (1990) 803.
\bibitem{hy1} M.B. Halpern and J.P. Yamron, Nucl. Phys. {\bf B332} (1990) 411.
\bibitem{wit2} E. Witten, in Memorial Volume for V. Knizhnik, ed. L. Brink et
             al., World Scientific, 1990.
\bibitem{irr} M.B. Halpern, E. Kiritsis, N.A. Obers, M. Porrati and J.P.
Yamron,
             Int. J. Mod. Phys. {\bf A5} (1990) 2275.
\bibitem{lie} M.B. Halpern, E.B. Kiritsis and N.A. Obers,
             Proceedings of the RIMS Research Project 1991,
              ``Infinite Analysis'', Int. J. Mod. Phys. A7,
              Suppl. {\bf 1A} (1992) 339.
\bibitem{rd} M.B. Halpern, ``{\em Recent Developments in the Virasoro Master
            Equation}'', in the proceedings of the Stony Brook conference,
            {\em Strings and Symmetries} 1991, World Scientific, 1992.
\bibitem{gt} M.B. Halpern and N.A. Obers, Comm. Math. Phys. {\bf 138} (1991)
63.
\bibitem{slg} M.B. Halpern and N.A. Obers, Int. J. Mod. Phys. {\bf A6} (1991)
              1835.
\bibitem{st} S. Schrans and W. Troost, Nucl. Phys. {\bf B345} (1990) 584.
\bibitem{hl} M.B. Halpern and N.A. Obers, Nucl. Phys. {\bf B345} (1990) 607.
\bibitem{mb} M.B. Halpern and N.A. Obers, Ann. of Phys. {\bf 212} (1991) 28.
\bibitem{nsc} M.B. Halpern and N.A. Obers, Int. J. Mod. Phys. {\bf A7} (1992)
              7263.
\bibitem{ssc} M.B. Halpern and N.A. Obers, Int. J. Mod. Phys. {\bf A7} (1992)
             3065.
\bibitem{sin} M.B. Halpern and N.A. Obers, J. Math. Phys. {\bf 32} (1991) 3231.
\bibitem{ggt} M.B. Halpern and N.A. Obers, J. Math. Phys. {\bf 33} (1992) 3274.
\bibitem{ks} Y. Kazama and H. Suzuki, Mod. Phys. Lett. {\bf A4} (1989) 235.
\bibitem{gep} R. Cohen and D. Gepner, Mod. Phys. Lett. {\bf A6} (1991) 2249.
\bibitem{hy2} M.B. Halpern and J.P. Yamron, Nucl. Phys. {\bf B351} (1991) 333.
\bibitem{sme} A. Giveon, M.B. Halpern, E.B. Kiritsis and N.A. Obers,
              Int. J. Mod. Phys. {\bf A7} (1992) 947.
\bibitem{jos} J.M. Figueroa-O'Farrill, ``{\em Constructing N=2 Superconformal
              Algebras out of N=1 Affine Lie Algebras}'', Bonn preprint,
              BONN-HE-93/21.
\bibitem{cf} A. Giveon, M.B. Halpern, E.B. Kiritsis and N.A. Obers,
              Nucl. Phys. {\bf B357} (1991) 655.
\bibitem{zam} A.B. Zamolodchikov, JETP Lett. {\bf 43} (1986) 730.
\bibitem{bpz} A.A. Belavin, A.M. Polyakov and A.B. Zamolodchikov, Nucl. Phys.
             {\bf B241} (1985) 691.
\bibitem{ns} A. Neveu and J.H. Schwarz, Nucl. Phys. {\bf B31} (1971) 6.
\bibitem{w}  A.B. Zamolodchikov, TMF {\bf 99} (1985) 108;
            V.A. Fateev and A.B. Zamolodchikov, Nucl. Phys. {\bf B280} (1987)
            644; V.A. Fateev and S.I. Lykyanov, Int. J. Mod. Phys. {\bf A3}
            (1988) 507.
\bibitem{co} M.B. Halpern and N.A. Obers, Int. J. Mod. Phys. {\bf A9}
             (1994) 265.
\bibitem{da} M.B. Halpern, Ann. of Phys. {\bf 194} (1989) 247.
\bibitem{wi} M.B. Halpern and N.A. Obers, Int. J. Mod. Phys. {\bf A9}
              (1994) 419.
\bibitem{lin} J. de Boer, K. Clubok and M.B. Halpern, Int. J. Mod. Phys.
             {\bf A9} (1994) 2451.
\bibitem{flat} M.B. Halpern and N.A. Obers,
            ``{\em Flat Connections and Non-Local
            Conserved Quantities in Irrational Conformal Field Theory}'',
            Berkeley preprint, UCB-PTH-9/33, hep-th 9312050. To appear in
             J. Math. Phys.
\bibitem{do} M.R. Douglas, ``{\em $G/H$ Conformal Field Theory}'', Caltech
             preprint, CALT-68-1453, 1987, unpublished.
\bibitem{rp} M.B. Halpern, ``{\em Recent Progress in Irrational Conformal
             Field Theory}'', Berkeley preprint, UCB-PTH-93/25, hep-th 9309087,
              1993. To appear in the proceedings of the Berkeley conference,
             {\it Strings 1993}.
\bibitem{hs} M.B. Halpern and N. Sochen, ``{\em Flat Connections for
             Characters in Irrational Conformal Field Theory}'', Berkeley
              preprint UCB-PTH-94/15, hep-th/9406076. To appear in Int. J.
             Mod. Phys. A.
\bibitem{ber} D. Bernard, Nucl. Phys. {\bf B303} (1988) 77.
\end{thebibliography}
\end{document}